\documentclass[reprint,aps,prd,nofootinbib,a4paper,10pt,twocolumn,superscriptaddress]{revtex4-2}
\usepackage{hyperref} 
\usepackage{changes}
\usepackage[section]{placeins}
\usepackage[titletoc]{appendix}
\usepackage{graphicx,subfigure}
\usepackage{xcolor}
\usepackage{dsfont}
\usepackage{bm}
\usepackage{mathtools} 
\usepackage{amsfonts}
\usepackage{enumerate}
\DeclareMathAlphabet{\mathpzc}{OT1}{pzc}{m}{it}

\begin{document}


\title{Machine learning cosmic backreaction and its effects on observations}
\author{S. M. Koksbang} 
\email{koksbang@cp3.sdu.dk}
\affiliation{CP3-Origins, University of Southern Denmark, Campusvej 55, DK-5230 Odense M, Denmark}

\begin{abstract}
	Symbolic expressions for cosmic backreaction and mean redshift drift in a range of 2-region models in terms of average quantities are presented. The demonstration that these expressions can be obtained constitutes the opening of a new avenue towards understanding the effects of cosmic backreaction in our universe: With a symbolic expression for the redshift drift at hand, the redshift drift can be used to constrain cosmological parameters including the large-scale expansion rate and backreaction. In addition, by introducing symbolic expressions for cosmic backreaction, this quantity can be constrained with observations such as redshift-distance measures. 
\end{abstract}
\keywords{Redshift drift, relativistic cosmology, observational cosmology, cosmological simulations} 
	
\maketitle
	
\paragraph*{Introduction}
The observed redshift of an astronomical object generally changes with time. This effect is known as redshift drift \cite{Sandage,McVittie}. In the Friedmann-Lemaitre-Robertson-Walker (FLRW) limit the redshift drift, $\delta z$, is given by
\begin{align}
	\delta z = \delta t_0\left[ (1+z)H_0 - H(z) \right],
\end{align}
where $\delta t_0$ is the observation time.
\newline\indent
As seen, redshift drift represents a direct measurement of the cosmic expansion rate and is thus considered an important future observable. However, the equation written above is only valid for FLRW spacetimes. The real universe contains structures which may affect the large-scale/average dynamics of the Universe. This effect is known as cosmic backreaction and was introduced in \cite{fluid1} where it was shown that the large-scale dynamical equations of a general inhomogeneous dust universe can be written as ($c=1$ and dots indicate time derivatives)
\begin{align}\label{eq:Buchert}
3H_D^2 & = 8\pi G \rho_D - \frac{1}{2}R_D - \frac{1}{2}Q\\
3\frac{\ddot a_D}{a_D} & = -4\pi G\rho_D + Q.
\end{align}
These equations are known as the Buchert equations and are valid for irrotational dust spacetimes with spatial hypersurfaces orthogonal to the dust flow.
\newline\indent
Subscripted $D$'s indicate the averaging domain, required to be larger than the assumed homogeneity scale. The Buchert equations are obtained by defining scalar averaging as
\begin{align}\label{eq:average_definition}
s_D := \frac{\int_D s dV}{\int_D dV},
\end{align}
where $s$ is some scalar, and applying this scheme to the Hamiltonian constraint and Raychaudhuri equation (see \cite{fluid1}). The kinematical backreaction $Q:=2/3\left[\left(\theta^2 \right)_D -\left(\theta_D \right)^2  \right] - \left(\sigma_{\mu\nu}\sigma^{\mu\nu} \right)_D$ is obtained from the dust expansion rate, $\theta$, and its shear tensor $\sigma_{\mu\nu}$. The volume averaged scale factor, $a_D$, is computed as $a_D = \left(V/V_0 \right)^{1/3}$, where $V$ is the proper (Riemannian) volume of the spatial averaging domain. The spatially averaged spatial curvature, $R_D$, may deviate from being proportional to $a_D^{-2}$ (its behavior in the FLRW limit). Together with $Q$, this deviation is called \emph{cosmic backreaction}. While $H_D$ can be computed as $\dot a_D/a_D$, it is also identical to a third of the average of the local expansion rate: $H_D = \frac{1}{3}\theta_D$.
\newline\indent
Since cosmic backreaction affects the dynamics of the Universe, it may be involved in observational cosmological tensions such as the $H_0$-tension \cite{tension1,tension2} as studied in \cite{tension_bc}. In addition, backreaction can lead to large-scale accelerated expansion without local acceleration \cite{conjecture1, conjecture4, conjecture5, conjecture6}, implying that backreaction could in principle make dark energy obsolete or at least significantly affect our phenomenological understanding of it.
\newline\indent
Whether or not backreaction truly is a good contestant to explain the accelerated expansion and, indeed, be of any significance in our universe at all, has been heavily debated \cite{debate1, debate2, debate3, debate4, debate5,debate6,debate7}. The bottom line is that we do not know and we probably cannot hope to learn until we are able to construct realistic relativistic models with full-fledged nonlinear structure formation including virialization so that we can study how backreaction appears and behaves in these. Additionally, we must find a connection between backreaction and observations. This latter point is the topic of interest here.
\newline\newline
\begin{figure*}
	\centering
	\subfigure[]{
		\includegraphics[scale = 0.5]{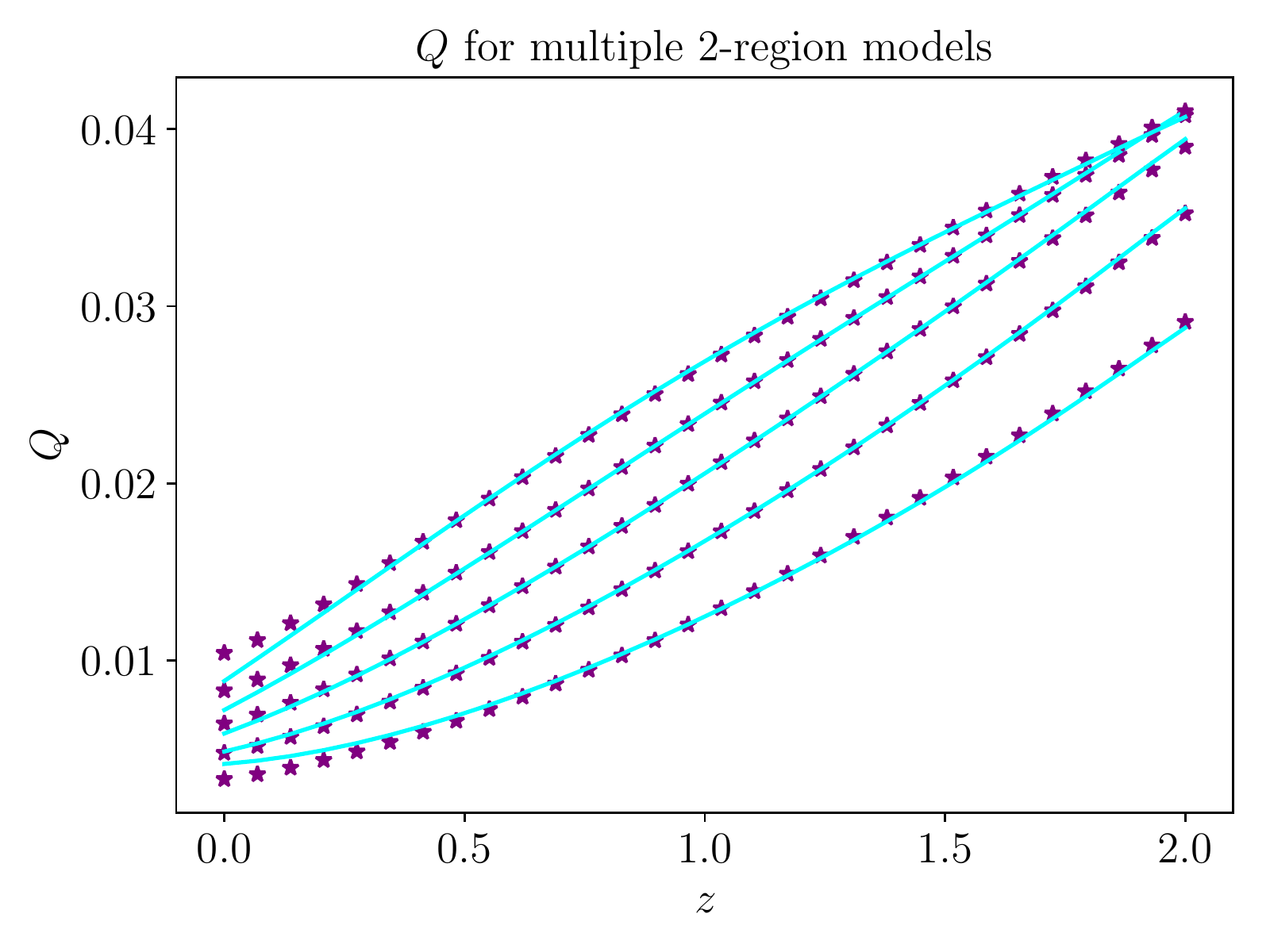}
	}
	\subfigure[]{
		\includegraphics[scale = 0.5]{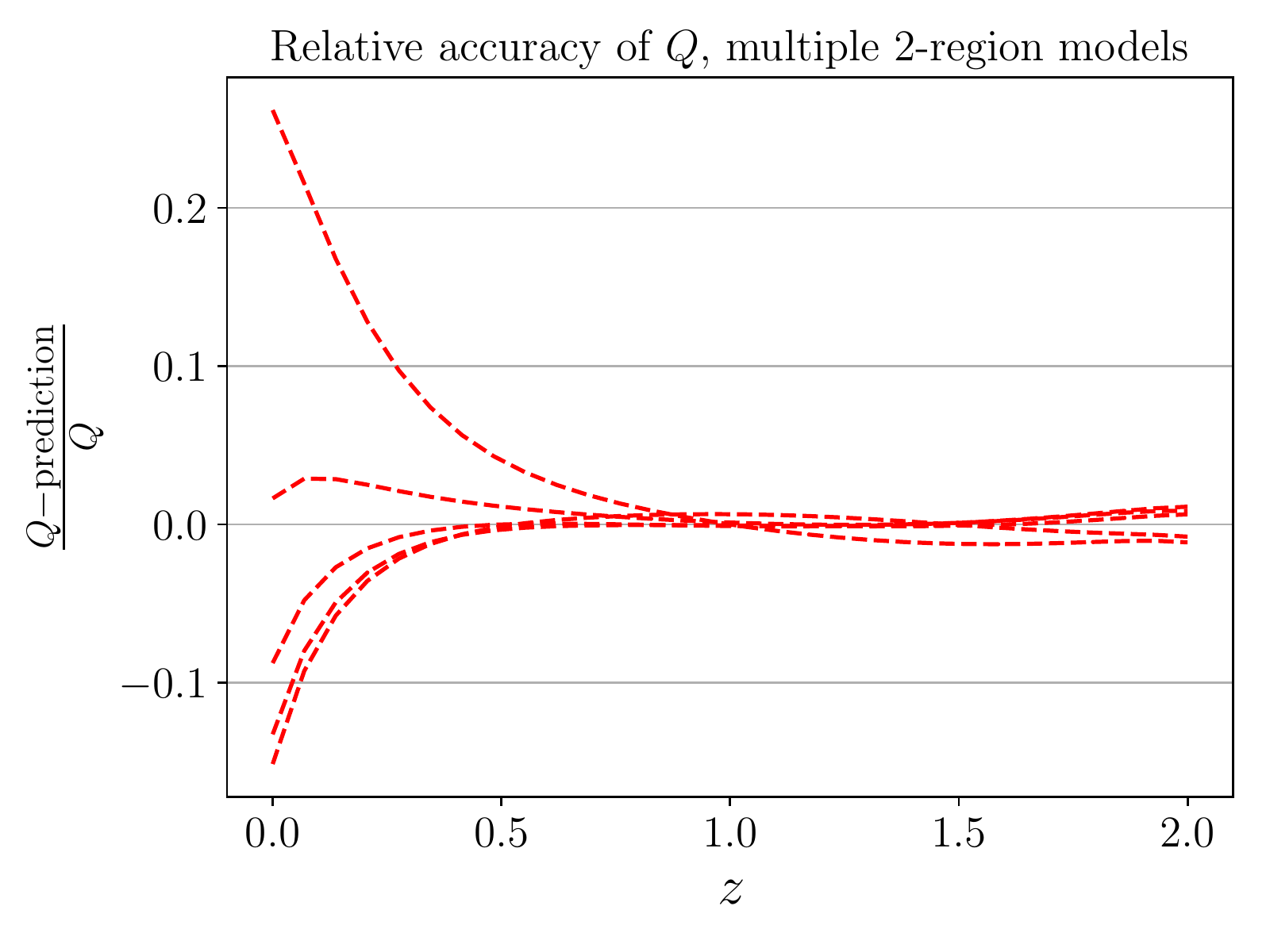}
	}
\par
	\centering
	\subfigure[]{
		\includegraphics[scale = 0.5]{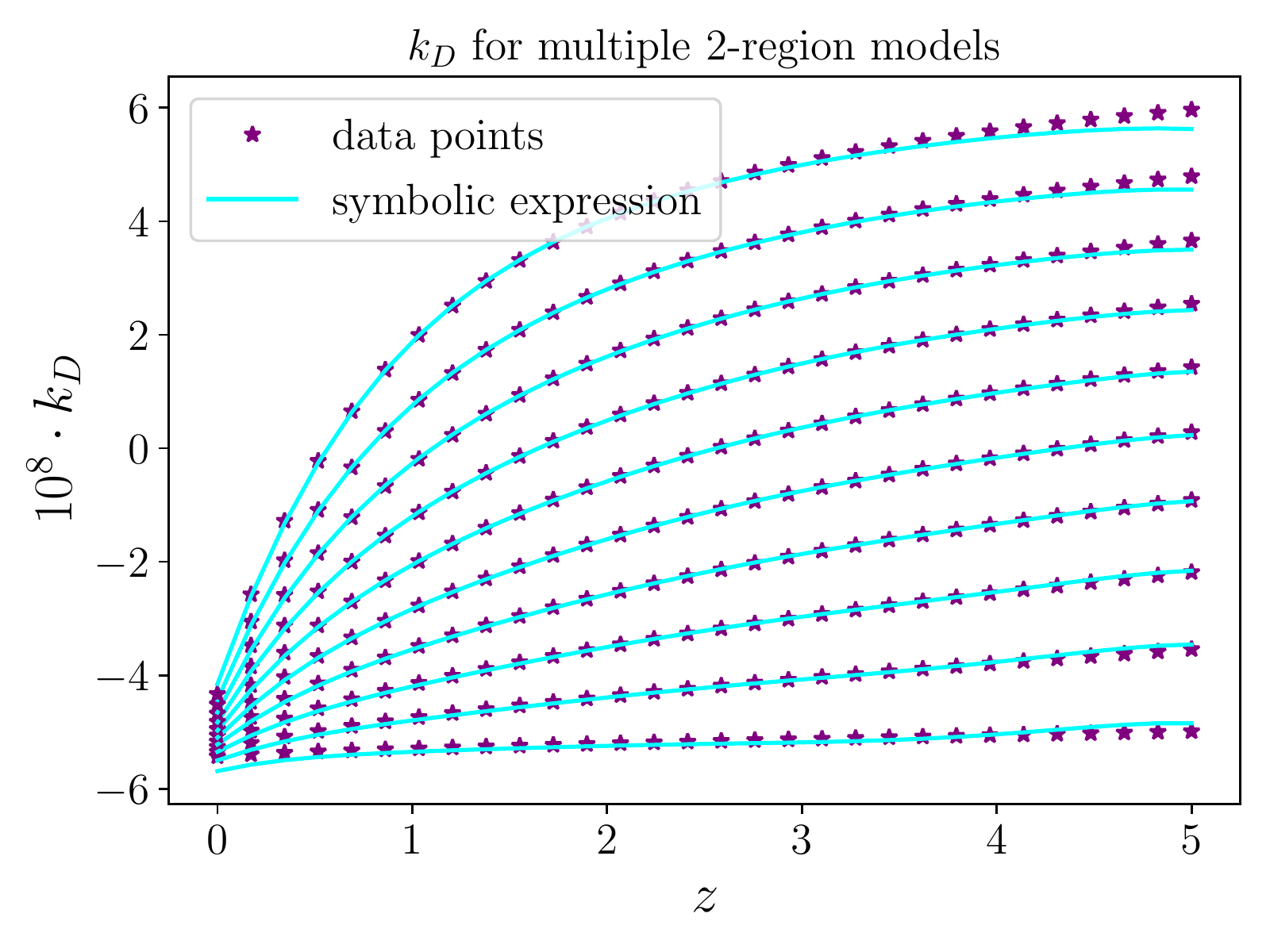}
	}
	\subfigure[]{
		\includegraphics[scale = 0.5]{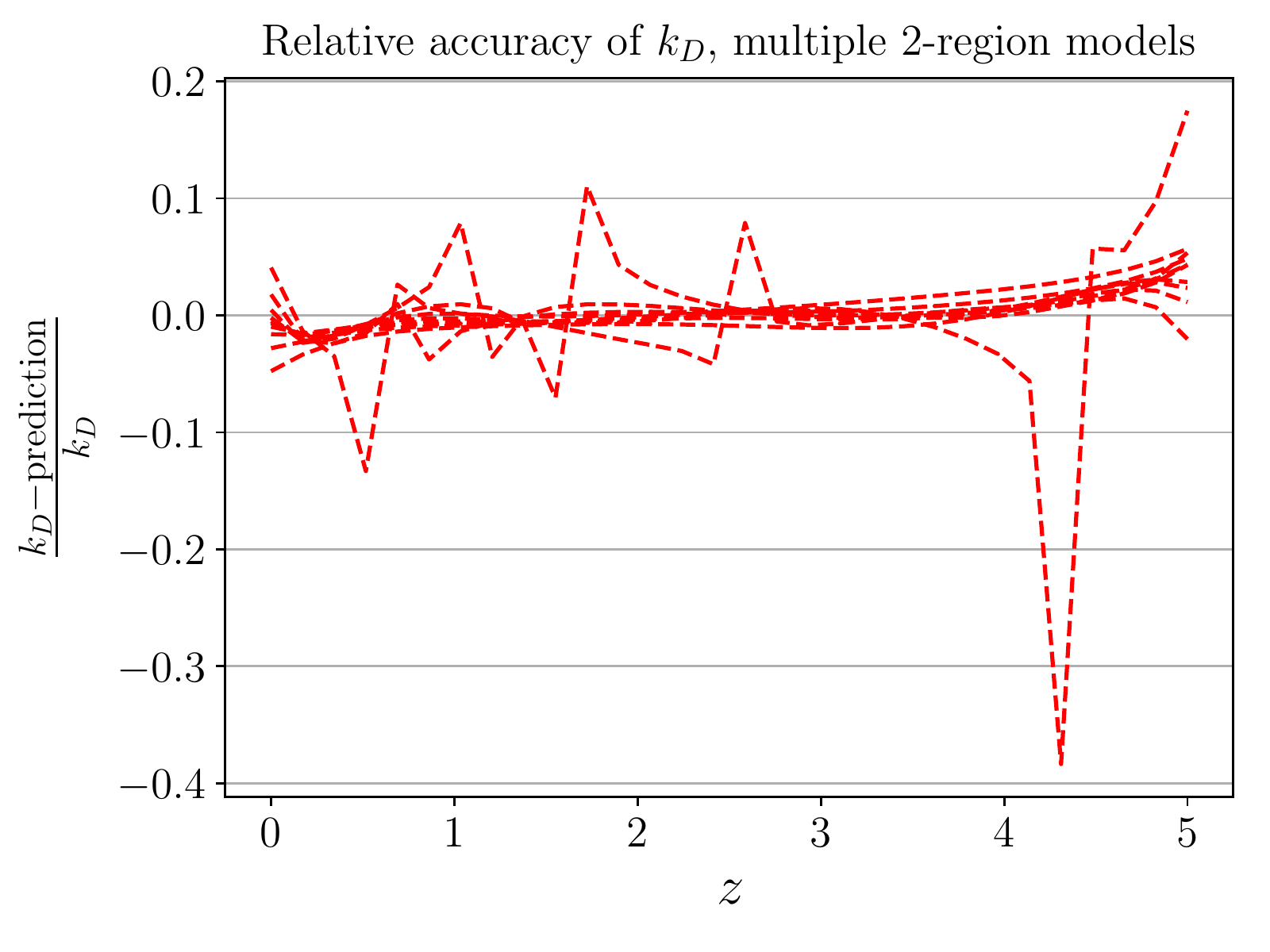}
	}	
	\caption{Symbolic expressions for $Q$ and $k_D$ together with data. The top part of this figure shows the same as part of figure 13 in \cite{joint_submission}.}
	\label{fig:backreaction}
\end{figure*}
\paragraph*{Observations in an inhomogeneous universe}
It has been shown \cite{syksy_light, another_look, Hellaby} that backreaction and spatially averaged quantities defined as in \cite{fluid1} can be related to the redshift-distance relation according to \cite{syksy_light}
\begin{align}
H_D\frac{d}{d\left\langle z\right\rangle }\left( (1+\left\langle z\right\rangle )H_D\frac{d\left\langle D_A\right\rangle }{d\left\langle z\right\rangle } \right) = -4\pi G\rho_D \left\langle D_A\right\rangle ,
\end{align}
where backreaction implicitly appears through $H_D$. Triangular brackets are used for denoting mean relations i.e. the mean redshift and redshift-distance relation obtained by averaging over several random lines of sight. With the above redshift-distance relation one ingredient is still missing before we can begin to sensibly constrain backreaction: We need to parameterize backreaction in terms of the volume averaged scale factor (or, equivalently, the mean redshift $\left\langle z\right\rangle = 1/a_D-1$ \cite{syksy_light}). In addition, it is not all observables that can be as simply related to spatial averages as the redshift-distance relation can. It has for instance been shown \cite{another_look, Hellaby, Asta1, Asta2} that the mean redshift drift is not equal to the drift of the mean redshift in a general inhomogeneous universe, even if spatial averages can be computed on hypersurfaces with statistical homogeneity and isotropy -- i.e. $\left\langle \delta z\right\rangle\neq \delta \left\langle z \right\rangle$. This means that, for a general spacetime, we cannot currently write a relation between spatial averages and the mean redshift drift.
\newline\newline
A new avenue towards overcoming these obstacles is introduced here where toy-models are used in an initial study into the possibility of using symbolic regression to obtain phenomenological expressions for the redshift drift and cosmic backreaction in terms of spatially averaged quantities.
\paragraph*{2-region models}
For simplicity, the task of obtaining symbolic expressions for backreaction and redshift drift is in this initial study done for 2-region models which are simple toy-models for which it is easy to generate backreaction and redshift drift data. 2-region models are disjoint ensembles of two different FLRW regions. It is straightforward to compute the spatial averages in such models \cite{2region_first2} and light propagation can be studied sensibly by arranging the two FLRW regions sequentially so that light rays propagate consecutively through each type of region. For such a model, the redshift drift along a light ray can be computed by simultaneously solving
\begin{align}
\frac{dt}{dr} &= -a\\
\frac{dz}{dr}& = (1+z)\dot a\\
\frac{d\delta z}{dr} &= \dot a\delta z + (1+z)\ddot a\delta t\\
\frac{d\delta t}{dr} &=-\dot a	\delta t,
\end{align}
where $\delta t$ is the difference in emission time of the two redshift signals. For results presented here, the initial condition for $\delta t$ was chosen as $\delta t_0 = 30$ years. This is consistent with typical values used in the literature but note that the precise value has no impact on the results presented here since the main effect of $\delta t_0$ is a scaling of $\delta z$. In the set of ODEs shown above, the scale factor, $a$, is to be evaluated according to the particular FLRW region the light ray is propagating through.
\newline\indent
Here, the two FLRW regions are chosen to be an empty FLRW region and an overdense matter+curvature FLRW region. The time coordinate of the models is parameterized in terms of a development angle, $\phi$, introduced through the standard parametric description of the matter+curvature FLRW model. Present time is chosen according to $\phi_0 = 3/2\pi$. We then have
\begin{align}
t& = t_0\frac{\phi-\sin(\phi)}{\phi_0-\sin(\phi_0)}\\
a_u& = \frac{f_u^{1/3}}{\pi}(\phi-\sin(\phi))\\
a_o & = \frac{f_o^{1/3}}{2}(1-\cos(\phi)),
\end{align}
where $a_o$ and $a_u$ are the scale factors of the over- and underdense regions, respectively. Different 2-region models are parameterized by $f:=f_o$ which represents the volume fraction of the overdense region at $\phi = \pi$. For details on the studied model, the reader is referred to the accompanying paper \cite{joint_submission}.
\begin{figure}
	\centering
	\includegraphics[scale = 0.5]{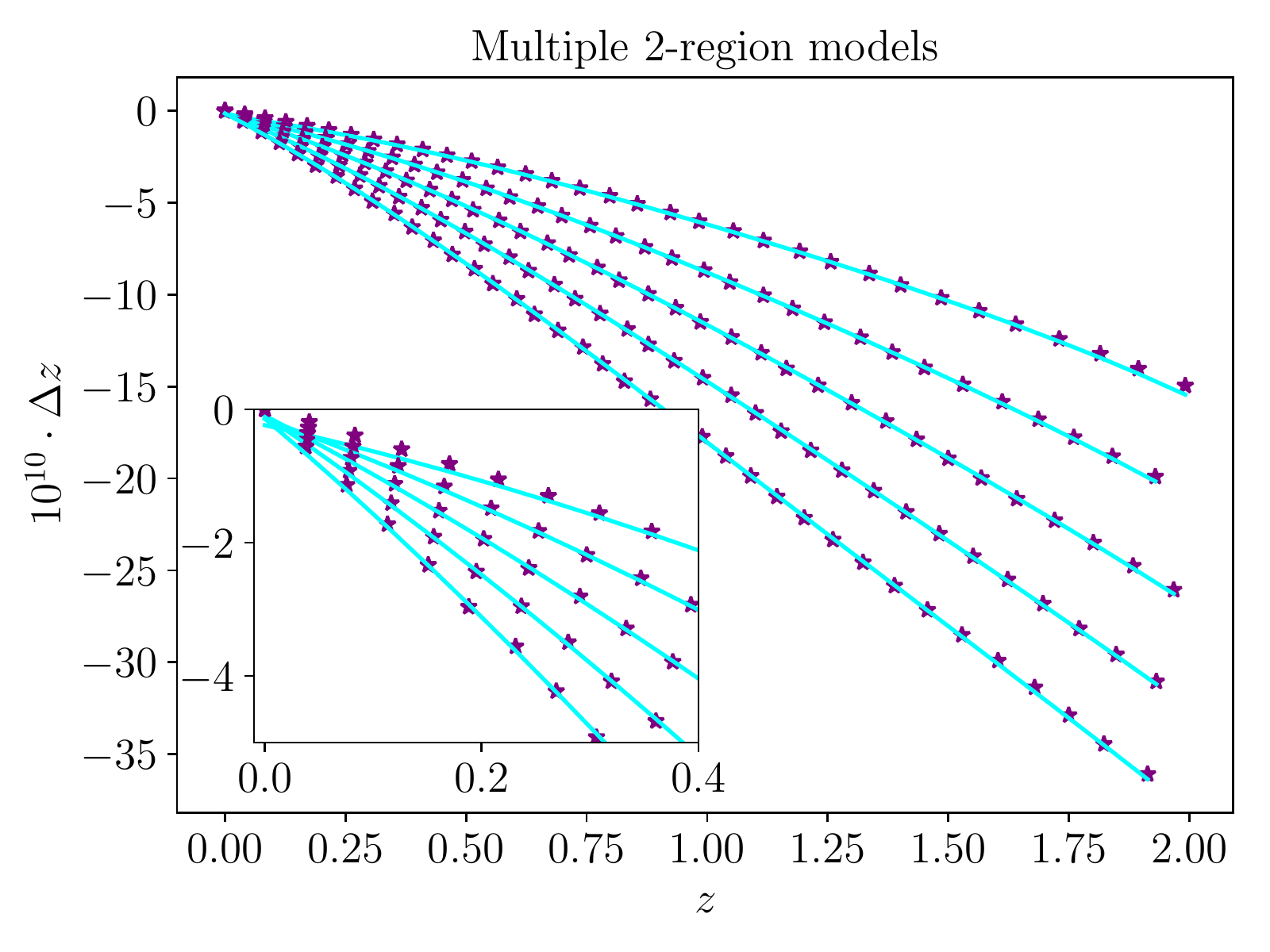}
	\caption{Symbolic expression (solid line) for $(z, f, 10^{10}\cdot \Delta z)$ together with data points (stars). A close-up is included to show the lack of accuracy at low values of the redshift.}
	\label{fig:Dz_multi}
\end{figure}

\paragraph*{Regression}
Symbolic expressions for redshift drift, $Q$ and $k_D:=R_D/a_D^2$ were obtained by using AI Feynman \cite{AIFeynman_1, AIFeynman_2}, a publicly available code for symbolic regression. The goal with symbolic regression is to discover symbolic expressions describing given data sets. While this can in principle be done solely using human skills\footnote{A famous examples of this, pointed out in \cite{AIFeynman_1}, is when Johannes Kepler spent 4 years doing ``symbolic regression'' to identify a mathematical description of Mars' orbit about the Sun.}, symbolic regression is here used to mean the automated approach where machine learning algorithms based on e.g. artificial neural networks (as in the case of AI Feynman) are utilized.
\newline\newline
To obtain a symbolic expression for $Q$, data of the form $(z,f,100\cdot Q)$ was presented to AI Feynman. The data was generated with $z\in[0,2]$ and $f\in[0.1,0.25]$ (see \cite{joint_submission} for a discussion of the choice of parameter ranges and the scaling of $Q$). Several runs with different choices of parameters for AI Feynman yielded no expressions with high (sub-percent) accuracy. The most accurate expression that was found is ($f_1+f_3+f_2$ in the notation of \cite{joint_submission})
\begin{align}
\nonumber 100\cdot Q &= 3.58069\cdot\tan^{-1}\left( (f\cdot(z+\exp(z)))\right) \\\nonumber
&  -0.00002\cdot \exp(\exp(z))+ 0.089282\cdot z^4 + 1.19043\cdot z^3f\\\nonumber
&- 0.51380\cdot z^3  - 15.87238\cdot z^2f^2 + 0.60670\cdot z^2f\\\nonumber & + 0.82042\cdot z^2 + 4.19790\cdot zf^2+ 2.86470\cdot zf \\\nonumber &- 0.83058z+ 12\cdot f^2 +11.67448\cdot f^2 - 4.43136\cdot f \\&+ 0.38435,
\end{align}
where the first term encapsulates the main trends of the data while the second term has minimal significance. A plot of the expression is shown in figure \ref{fig:backreaction} together with data points. Figure \ref{fig:backreaction} also shows the relative deviation between the data points and the symbolic expression. As seen, the symbolic expression is not very accurate at low redshifts but from $z\approx 0.5$ the expression is accurate at percent level for most of the considered $f$-interval.
\newline\indent
Once a symbolic expression for $Q$ is known, $R_D$ can be obtained through the integrability condition 
\begin{align}\label{eq:constraint}
a_D^{-6} \left(a_D^6 Q \right)^.  + a_D^{-2} \left(a_D^2 R_D  \right)^. = 0,
\end{align}
which must be fulfilled in order for the two Buchert equations to be consistent with each other. The opposite is also true: One could do symbolic regression to obtain an expression for $R_D$ and use the integrability condition to obtain $Q$. Which procedure is the most useful depends on which quantity AI Feynman obtains the most accurate description for. For the models studied here, it turned out that the expressions obtained for $R_D$ are more accurate than those obtained for $Q$. In addition, accurate expressions could be obtained on larger feature intervals. This is seen in figure \ref{fig:backreaction}, where the subfigures in the bottom row show the most accurate expression obtained for $k_D:=R_Da_D^2$ using data on the feature intervals $z\in[0,5]$ and $f\in[0.01,0.3]$. Note that the relative accuracy depicted for $k_D$ has spikes. These come from division by zero and are not due to large inaccuracy. The symbolic expression depicted in figure \ref{fig:backreaction} for $(z,f,10^8k_D)$ is
\begin{align}
\nonumber 10^8\cdot k_D &=\\\nonumber & -0.00219 \cdot z^6 + 0.03768\cdot z^5 f + 0.03168\cdot z^5\\\nonumber & - 0.64965\cdot z^4 f^2 - 0.43309\cdot z^4 f - 0.17684\cdot z^4\\\nonumber &  + 8.84545\cdot z^3 f^2 + 1.52360\cdot z^3 f+ 0.49810\cdot z^3\\\nonumber &  + 8.09975\cdot z^2f^3 - 45.06587\cdot z^2f^2 - 1.95429\cdot z^2f\\\nonumber & - 0.78006\cdot z^2 - 40.49877\cdot zf^3 + 80.42749\cdot zf^2\\\nonumber & + 9.97251\cdot zf + 0.67266\cdot z + 67.56565\cdot f^3\\ & - 25.90655\cdot f^2 + 7.02894\cdot f - 5.75576.
\end{align}
The expression is only accurate inside the feature intervals used for generating data for AI Feynman and quickly becomes highly inaccurate outside this feature region.
\\\\
\begin{figure}
	\centering
	\includegraphics[scale = 0.5]{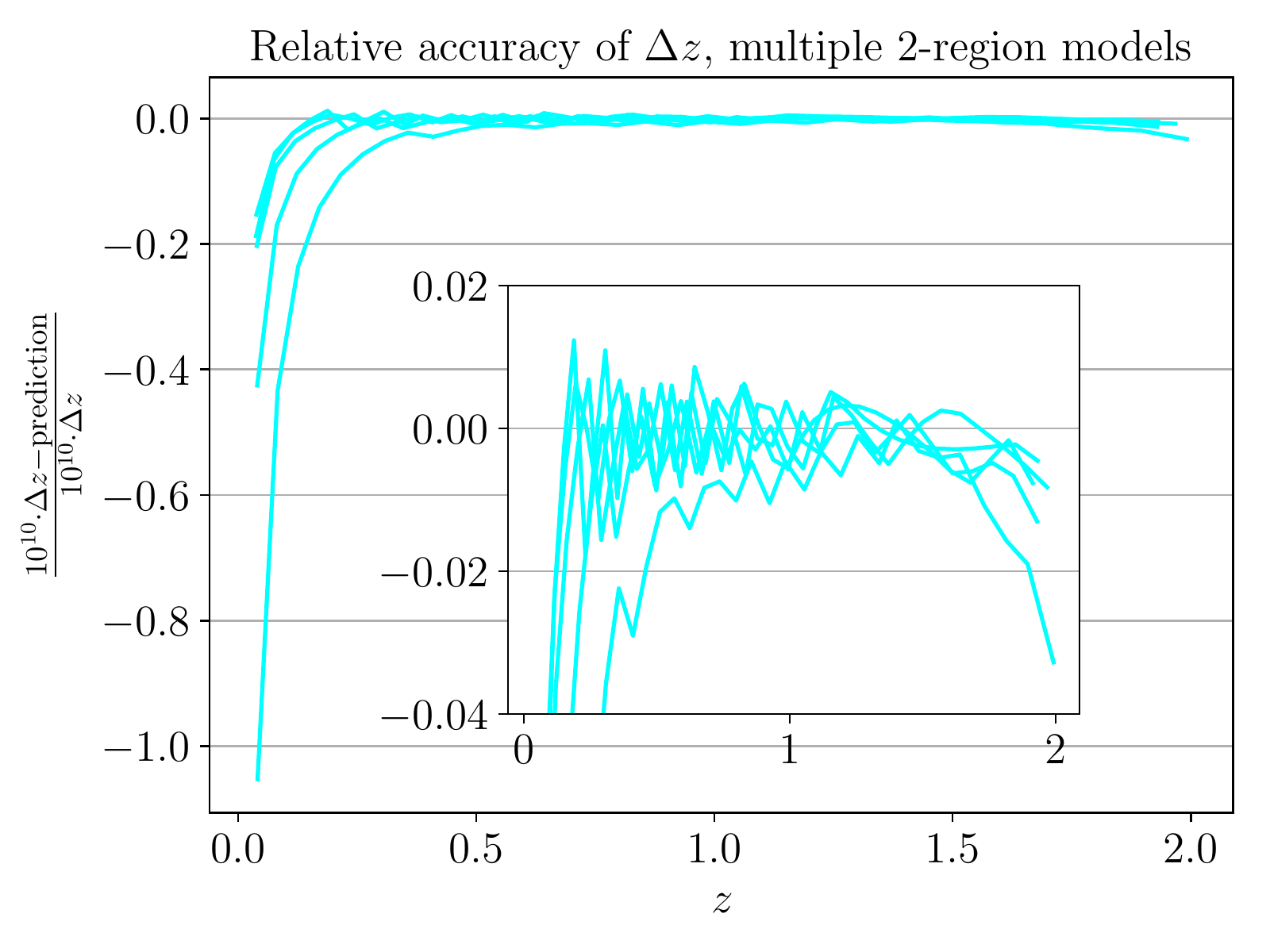}
	\caption{Relative accuracy of the symbolic expression for $(z, f, 10^{10}\cdot \Delta z)$. This figure is the same as one of the subfigures in figure 16 of \cite{joint_submission}. }
	\label{fig:Dz_rel}
\end{figure}
A symbolic expression for the mean redshift drift was obtained with data in the form $(z,f,10^{10}\cdot \Delta z)$. The quantity $\Delta z$ is defined as the difference between the mean redshift drift and the drift of the mean redshift. Remembering that triangular brackets indicate taking the mean over many random observations/lines of sight, we can write $\Delta z :=<\delta z>-\delta<z>$, where
\begin{align}\label{eq:dz_naive}
\delta <z> = \delta t_0\left[ (1+z)H_{D_0} - H_D \right] .
\end{align}
By presenting this to the AI Feynman algorithm, several fairly accurate (percent-level) symbolic expressions were obtained, although only by considering the smaller feature interval also used for the kinematical backreaction, i.e. $z\in[0,2]$ and $f\in[0.1,0.25]$. The expression depicted in figure \ref{fig:Dz_multi} is the polynomial
\begin{align}
\nonumber F(z,f) & =10^{10}\Delta  z=\\
&\nonumber -10^{-5}\cdot z^6 - 0.05383\cdot z^5 f - 0.16136\cdot z^5 \\\nonumber &- 51.42798\cdot z^4f^2 + 16.24516\cdot z^4f - 0.71557\cdot z^4 \\\nonumber &- 1.41156\cdot z^3f^3 + 207.32051\cdot z^3 f^2 - 58.40215\cdot z^3 f\\\nonumber & + 4.05026\cdot z^3 - 0.01453\cdot z^2f^4
+ 4.26328\cdot z^2f^3 \\\nonumber &- 191.11012\cdot z^2f^2 + 41.45695\cdot z^2 f - 5.05871\cdot z^2 \\\nonumber &-7\cdot 10^{-5}\cdot zf^5 + 0.02924\cdot z f^4 - 2.61453\cdot zf^3\\\nonumber & - 30.04826\cdot zf^2 - 56.13623\cdot zf + 2.28806\cdot z \\\nonumber+ & 7\cdot 10^{-5}\cdot f^5 - 0.00896\cdot f^4 - 0.20814\cdot f^3\\ & - 16.98247\cdot f^2 + 6.61763\cdot f - 0.72982,
\end{align}
but it is noted that other, simpler expressions were also found to be accurate on major parts of the studied feature region. The choice of prioritizing accuracy over simplicity regarding which expressions to show is based on an expectation that accuracy is more important for parameter constraints than simplicity. It is unclear from the current literature to what extent this is true and this expectation is therefore the topic of an ongoing study.
\newline\indent
As shown in figure 16 in \cite{joint_submission}, the expression found above is also accurate when extrapolated somewhat outside the interval of $f$ used for generating data for AI Feynman. The expression does {\em not} extrapolate well to values of the redshift above 2.
\newline\indent
Figure \ref{fig:Dz_multi} includes a close-up of the low-$z$ region to show that the symbolic expression is inaccurate at low vales of the redshift, especially for the smaller values of $f$.  This is also seen in figure \ref{fig:Dz_rel} which shows the relative accuracy of the symbolic expression. As seen, the symbolic expression is accurate within a few percent for the major part of the studied feature region and is even sub-percent on part of it. The poor accuracy at low redshift values is not considered too discouraging since the data presented to AI Feynman is actually local redshift and redshift drift rather than mean values. While the local redshift and redshift drift computations coincide closely with the mean values in the studied models, using the local values means that the data contains noise in the form of statistical fluctuations. These fluctuations are most prominent at low redshift and could be removed by taking the mean over several light rays with observers placed differently in the two types of FLRW regions.
\newline\indent
With $F(z,f)$ at hand, we can write the mean redshift drift as
\begin{align}
<\delta z> = \delta t_0\left[ (1+z)H_{D_0} - H_D \right] + 10^{-10}\cdot F(z,f).
\end{align}
\paragraph*{Discussion and conclusions} \label{sec:conclusion}
Expressions obtained with symbolic regression lack theoretical underpinning and must, as an onset, be considered phenomenological. The expressions can nonetheless be useful for guiding a theoretical understanding of under what circumstances backreaction becomes relevant and how realistic it is to obtain general parameterizations of the quantity in terms of e.g. the mean redshift. This can e.g. be done by focusing studies on obtaining symbolic expressions that extrapolate well outside the parameter regions used for obtaining the expressions. Similarly, expressions obtained for redshift drift can be used to guide theoretical studies into the circumstances required for the mean redshift drift and drift of the mean redshift to be similar.
\newline\indent
Even without theoretical underpinning, it is valid to use the symbolic expressions together with data to constrain model parameters. However, this must not be done blindly with \emph{any} result obtained using machine learning but must instead be done in combination with considerations of the theoretical setup which can e.g. inform on the relevance of combining different features and targets. The symbolic expressions presented here are only valid for 2-region models in the studied model parameter and redshift intervals. By applying the method to more realistic models -- the topic of future work -- the presented approach constitutes a new avenue within the field of \emph{inhomogeneous cosmology}, with the goal to connect backreaction to observables in a physically motivated manner. In relation to this it may be insightful to study generalizations of the simple 2-region models considered here such as the multi-scale models of \cite{multiscale1,multiscale2} or the more sophisticated timescape version \cite{simple_timescape}.

\vspace{6pt} 
\begin{acknowledgments}
	The author thanks the anonymous referees for their comments which have significantly improved the presentation of the work. During the final stages of the review process, the author transitioned from being funded by the Carlsberg Foundation to being funded by VILLUM FONDEN, grant VIL53032.
\end{acknowledgments}

\end{document}